\documentclass[pre,twocolumn,showpacs]{revtex4-1}
\usepackage{color,graphicx,hyperref,amsmath,ulem}
\newcommand{\im}{\mathrm{i}}
\begin{document}
\title{The Kuramoto model in presence of additional interactions that break
rotational symmetry}
\author{V K Chandrasekar$^1$, M Manoranjani$^1$ and Shamik Gupta$^{2,3}$}
\affiliation{$^1$Centre for Nonlinear Science \& Engineering, School of
Electrical \& Electronics Engineering, SASTRA Deemed University,
Thanjavur-613 401, Tamil Nadu, India \\ $^2$Department of Physics, Ramakrishna Mission Vivekananda
Educational and Research Institute, Belur Math, Howrah 711202, India \\
$^3$Regular Associate, Quantitative Life Sciences Section, ICTP - The Abdus Salam International Centre for Theoretical Physics, Strada Costiera 11, 34151 Trieste, Italy}
\date{\today}
\begin{abstract}
The Kuramoto model serves as a paradigm to study the phenomenon of
spontaneous collective synchronization. 
We study here a nontrivial generalization of the Kuramoto model by
including an interaction that breaks explicitly the rotational symmetry
of the model. In an inertial frame (e.g., the laboratory frame), the Kuramoto model does not allow for a stationary
        state, that is, a state with time-independent value of the so-called Kuramoto
        (complex) synchronization order parameter $z\equiv re^{i\psi}$;
        Note that a
time-independent $z$ implies $r$ and $\psi$ both time independent, with
the latter fact corresponding to a state in which $\psi$ rotates at zero
        frequency (no rotation). In this backdrop, we ask: Does the introduction of
the symmetry-breaking term suffice to allow for the existence of a stationary state in the laboratory frame? 
Compared to the original model, we reveal a rather rich phase diagram of the resulting
model, with the existence of both stationary and standing wave
phases. While in the former the synchronization order parameter $r$ has a
long-time value that is time independent, one has in the latter an
oscillatory behavior of the order parameter as a function of time that
nevertheless yields a non-zero and time-independent time average. Our
results are based on numerical integration of the dynamical equations as
well as an exact analysis of the dynamics by invoking the so-called Ott-Antonsen ansatz that allows to derive a reduced set of
time-evolution equations for the order parameter.
\end{abstract}
\maketitle
\section{Introduction: Model and Summary of Results}
\label{sec:intro}

One of the widely invoked and extensively studied models of spontaneous
collective synchronization in the field of nonlinear dynamics is the
so-called Kuramoto model~\cite{Pikovsky:2001}. The setting of the model allows to apply it to
study a wide range of physical systems pervading length and time scales
of several order of magnitude, ranging from groups of fireflies
flashing on and off in unison~\cite{Buck:1988}, cardiac pacemaker
cells~\cite{Peskin:1975}, electrochemical~\cite{Kiss:2002}
and electronic~\cite{Temirbayev:2012} oscillators, to Josephson junction
arrays~\cite{Benz:1991},
audience clapping in unison~\cite{Zeda:2000}, electrical power-grid
networks~\cite{Rohden:2012}, in discussing adaptive networks in
neuroscience and social sciences~\cite{Scholl}, etc. The model comprises
nearly-identical $N$ limit-cycle
oscillators with distributed natural frequencies
$\omega_j;~j=1,2,\ldots,N$, interacting weakly with one another, with the strength of coupling
being the same for every pair of oscillators~\cite{Kuramoto:1984,Strogatz:2000,Acebron:2005,Gupta:2014,Gupta:2018}. Denoting by $\theta_j \in
[-\pi, \pi]$ the phase of the $j$-th oscillator, the dynamics of the model
is described by a set
of $N$ coupled first-order nonlinear differential equations of the form
\begin{equation}
\frac{{\rm d}\theta_j}{{\rm d}t}=\omega_j+\frac{K}{N}\sum_{k=1}^N
\sin(\theta_k-\theta_j),
\label{eq:eomKM}
\end{equation}
where $K \ge 0$ is the coupling constant. The frequencies
$\{\omega_j\}_{1\le j \le N}$ denote a set of
quenched-disordered random variables distributed
according to a common distribution $g(\omega)$, with normalization
$\int_{-\infty}^\infty{\rm d}\omega~g(\omega)=1$ and finite mean
$\omega_0$. 
The Kuramoto synchronization order
parameter, giving a measure of synchrony present in the
system at time $t$, is defined as~\cite{Strogatz:2000}
\begin{eqnarray}
&&z(t)=r(t)e^{\im\psi(t)}\equiv \frac{1}{N}\sum_{j=1}^N
        e^{\im\theta_j(t)}=r_x(t)+\im r_y(t); \label{eq:z}\\
&&
(r_x,r_y)(t)\equiv \frac{1}{N}\sum_{j=1}^N \left(\cos \theta_j,\sin
\theta_j\right)(t).
\label{eq:r}
\end{eqnarray}
The quantity $r(t);~0 \le r(t) \le 1$, measures the amount
of synchrony present in the system at time $t$, while
$\psi(t)=\tan^{-1}(r_y(t)/r_x(t))$ gives the average phase. When the
oscillators are incoherent or unsynchronized so that over a stretch of
time or in an ensemble of $\theta_j$-values at a given time, one
has with equal probabilities $e^{\im\theta_j}$ for any $j$ equal to any complex number with
modulus unity, $r(t)$ averages
to zero. On the other hand, $r(t)$ has a non-zero average in the
synchronized phase in which a finite
fraction of oscillators have phase differences that are constant in time.

The dynamics~(\ref{eq:eomKM}) satisfies rotational symmetry, whereby rotating every phase by an
arbitrary angle same for all leaves the dynamics invariant. In
particular, one may implement the transformation
$\theta_j(t) \to \theta_j(t)+\omega_0 t~\forall~j,t$, which is
tantamount to viewing the dynamics in a frame rotating uniformly with
frequency $\omega_0$ with respect to an inertial frame, e.g., the
laboratory frame. In such a
comoving frame, the frequencies $\omega_j$ follow the shifted distribution $g(\omega+\omega_0)$, thereby having zero mean.

The model~(\ref{eq:eomKM}) has been extensively studied over the years and a
host of results have been derived, see
Ref.~\cite{Gupta:2018} for a recent overview. For example, consider a $g(\omega)$ that is unimodal, namely, a $g(\omega)$ that is symmetric about its mean $\omega_0$ and decreases monotonically and
continuously to zero with increasing $|\omega - \omega_0|$.  Considering
the limit $N \to \infty$, it is then
known that in the stationary state of the dynamics~(\ref{eq:eomKM}), attained in the limit $t \to \infty$ and in the comoving frame, the system may exist in either a synchronized or an incoherent phase depending on whether the coupling $K$
is respectively above or below a critical threshold $K_c=2/(\pi
g(\omega+\omega_0))$. On tuning $K$ across $K_c$ from high to low
values, one observes a continuous phase transition in $r_{\rm st}$, the
stationary value of $r(t)$; Namely, $r_{\rm st}$ decreases continuously
from the value of unity, achieved as $K \to \infty$, to zero at $K=K_c$
and remains zero at smaller $K$ values. It is then usual to interpret the transition as
the case of a supercritical bifurcation, in which on tuning $K$ from low
to high values, a
synchronized phase bifurcates from the incoherent phase at $K=K_c$. In
particular, a small change of $K$ across $K_c$ results in only a small
change in the value of $r_{\rm st}$. The transition could also be of first
order (e.g., for a bimodal $g(\omega)$~\cite{Martens:2009} or in the inertial version of the
dynamics~(\ref{eq:eomKM}) \cite{Gupta:2014}), whereby $r_{\rm st}$ exhibits an abrupt and big change on changing $K$ by a small amount across the
phase transition point; in this case, the bifurcation is said to be subcritical and
leads to hysteresis~\cite{Gupta:2018}. For discussions on general form of
the Kuramoto model and arbitrary frequency distributions, see, e.g., Refs.~\cite{Aneta1,Aneta2}. 

In this work, we consider a generalization of the Kuramoto dynamics~(\ref{eq:eomKM})
by including an interaction term that explicitly breaks the rotational
symmetry of the dynamics. To this end, we consider the
following set of $N$ coupled nonlinear differential equations:
\begin{equation}
\frac{{\rm d}\theta_j}{{\rm d}t}=\omega_j+\frac{1}{N}\left[\epsilon_1
\sum_{k=1}^N \sin(\theta_k-\theta_j)+\epsilon_2\sum_{k=1}^N
\sin(\theta_k+\theta_j)\right],
\label{eq:eom}
\end{equation}
where the real parameters $\epsilon_{1,2}$ denote the coupling
constants. In terms of the quantities $r_x,r_y$,
Eq.~(\ref{eq:r}), the dynamics~(\ref{eq:eom}) reads 
\begin{equation}
\frac{{\rm d}\theta_j}{{\rm d}t}=\omega_j+(\epsilon_1+\epsilon_2) r_y
\cos \theta_j+ (\epsilon_2-\epsilon_1)r_x \sin \theta_j.
\label{eq:eom-1}
\end{equation}
The above equation makes it evident that the quantities $r_x$ and $r_y$
act as mean-fields determining the motion of every oscillator in the
ensemble. In Appendix~\ref{app1}, we motivate the form of the
dynamics~(\ref{eq:eom}) by considering the dynamics of a collection of $N$ globally-coupled Stuart-Landau
limit-cycle oscillators with conjugate feedback.

Note that setting $\epsilon_2=0$ in Eq.~(\ref{eq:eom}) (equivalently,
Eq.~(\ref{eq:eom-1})) reduces the dynamics to that of the Kuramoto model~(\ref{eq:eomKM}) on
identifying $\epsilon_1$ with the parameter $K \ge 0$. 
We therefore take $\epsilon_1$ to be positive. Then, for a fixed $\epsilon_1$,
changing $\epsilon_2$ to $-\epsilon_2$ is tantamount to keeping
$\epsilon_2$ unchanged but effecting the transformation $\theta_j \to
\theta_j + \pi/2~\forall~j$ in Eq.~(\ref{eq:eom}) (redefinition of the
origin with respect to which the phases are measured). Consequently, we
may take $\epsilon_2 \ge 0$, without loss of generality. Rotational symmetry is possible
in the dynamics~(\ref{eq:eom}) only with the choice $\epsilon_2=0$, so that the $\epsilon_2$-term in the dynamics may be
interpreted as a rotational-symmetry-breaking interaction. 
In contrast to the Kuramoto model, the dynamics~(\ref{eq:eom}) is not invariant
with respect to the transformation $\theta_j \to \theta_j' \equiv \theta_j + \omega_0
t~\forall~j,t$
because of the $\epsilon_2$-term. Indeed, under such a
transformation, we get
\begin{eqnarray}
\frac{{\rm d}\theta_j'}{{\rm d}t}&=&\omega_j+\omega_0+\frac{1}{N}\Big[\epsilon_1
\sum_{k=1}^N \sin(\theta_k'-\theta_j')\nonumber \\
&&+\epsilon_2\sum_{k=1}^N
\sin(\theta_k'+\theta_j'-2\omega_0t)\Big],
\label{eq:eom-transformed}
\end{eqnarray}
which does not have the same form as the dynamics~(\ref{eq:eom}), and so the
transformation does not leave the dynamics invariant. As a
result, the mean $\omega_0$ is expected to have an essential
effect on the dynamics~(\ref{eq:eom}), which cannot be gotten rid of by viewing the
dynamics in a frame rotating uniformly
with frequency $\omega_0$ with respect to the laboratory frame, as is
possible with the Kuramoto model. From now on, we will study the
dynamics~(\ref{eq:eom}) only in the inertial frame (i.e., the laboratory
frame) and not in the
comoving frame.

As is usual with studies of the Kuramoto model, we will consider in this
work a unimodal $g(\omega)$. Specifically, we will consider two
representative choices, namely, a Lorentzian:
\begin{equation}
g(\omega)=\frac{\gamma}{\pi((\omega-\omega_0)^2+\gamma^2)};~~\gamma >0,
\label{eq:lor}
\end{equation}
and a Gaussian:
\begin{equation}
g(\omega)=\frac{1}{\sqrt{2\pi
\sigma^2}}\exp(-(\omega-\omega_0)^2/(2\sigma^2));~~\sigma^2 >0.
\label{eq:gaussian}
\end{equation}

Let us remark on a relevant aspect of the
dynamics~(\ref{eq:eom}). Summing both sides of the equation over $j$, we
get in the limit $N \to \infty$ that the mean ensemble frequency of the
$\theta_j$'s is given by 
\begin{equation}
f \equiv \frac{{\rm d}}{{\rm d}t}\left(\frac{1}{N}\sum_{j=1}^N
\theta_j\right)=\omega_0+2 \epsilon_2 r_xr_y,
\label{eq:mean-ensemble-frequency}
\end{equation}
where we have used the fact that $\lim_{N\to \infty}(1/N)\sum_{j=1}^N
\omega_j=\omega_0$. From Eq.~(\ref{eq:mean-ensemble-frequency}), we see
that the mean ensemble frequency $f$ coincides with the mean of the natural
frequency distribution $g(\omega)$ when the dynamics~(\ref{eq:eom}) becomes
that of the Kuramoto model (that is, $\epsilon_2=0$). In our case, with
$\epsilon_2 \ne 0$, the two frequencies would in general not coincide
unless the system is unsynchronized so that $r_x=r_y=0$.

Considering the limit $N \to \infty$, this work aims at a detailed
characterization of the long-time ($t\to \infty$) limit of the
dynamics~(\ref{eq:eom}), equivalently, Eq.~(\ref{eq:eom-1}), with the primary objective of identifying and
understanding what new features are brought in by the
introduction of the rotational-symmetry-breaking $\epsilon_2$-term.
Now, in the absence of the $\epsilon_2$-term,
the dynamics does not allow for a stationary state in the laboratory
frame. Namely, one cannot have in such a frame time-independent $z$: in the synchronized phase, $\psi$
will change uniformly in time with frequency $\omega_0$. Note that a
time-independent $z$ implies $r$ and $\psi$ both time independent, with
the latter fact corresponding to a state in which $\psi$ rotates at zero
        frequency (no rotation). In this
backdrop, we ask: Does the introduction of the $\epsilon_2$-term suffice to
allow for the existence of a stationary state in the laboratory frame?
If the answer is in the affirmative, some immediate and pertinent
questions follow: What is
the nature of the stationary state? Is there a range of parameter values
for which one has a synchronized stationary state? What is the complete phase diagram
in the ($\epsilon_1-\epsilon_2)$-plane? Can one characterize the phase
diagram analytically?

\begin{figure}[!ht]
\centering
\includegraphics[width=8cm]{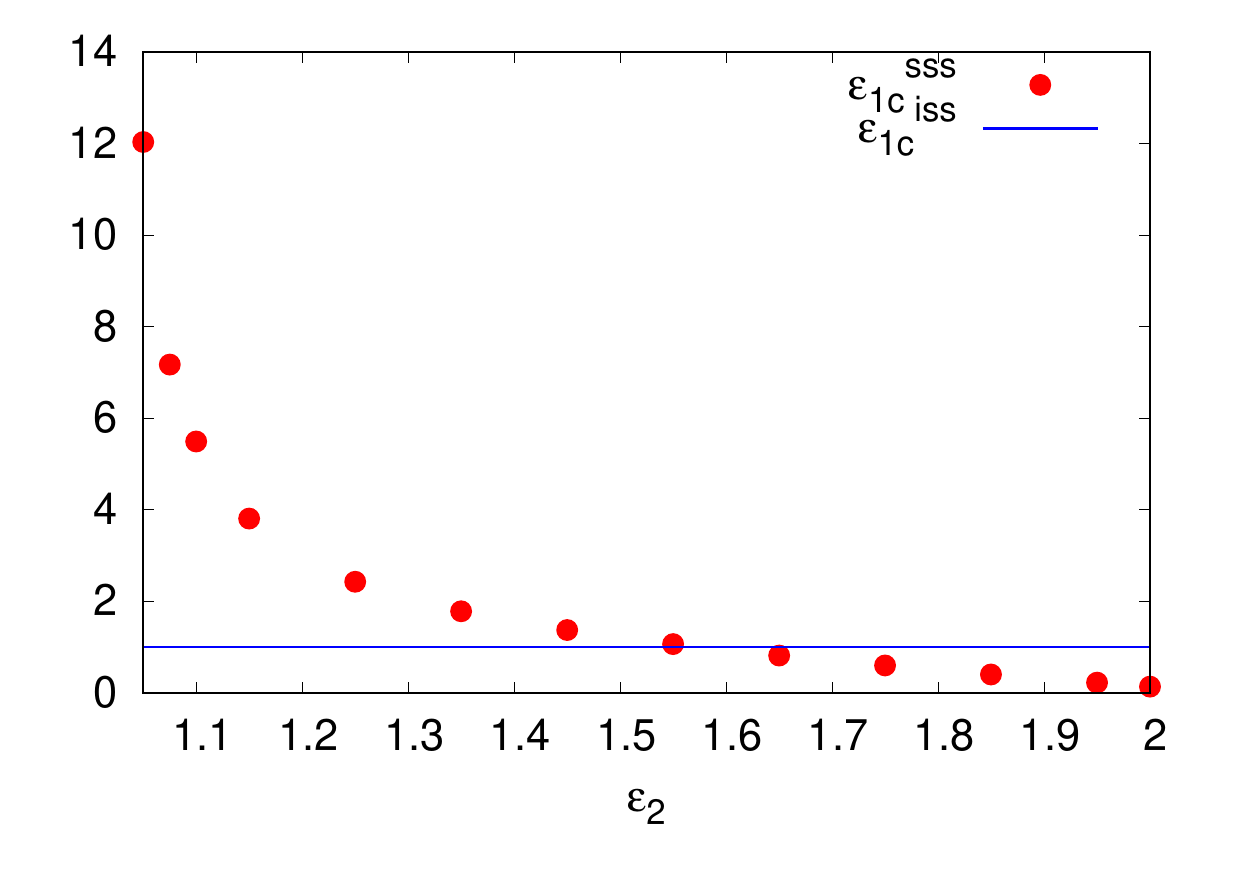}
\caption{For the model~(\ref{eq:eom}) and considering for
$\omega_j$'s the Lorentzian distribution~(\ref{eq:lor}) with $\gamma=0.5$ and $\omega_0=1.0$, the figure shows the thresholds $\epsilon_{1c}^{\rm ISS}(\epsilon_2)$ and $\epsilon_{1c}^{\rm
SSS}(\epsilon_2)$ as a
function of $\epsilon_2$, where $\epsilon_{1c}^{\rm
ISS}(\epsilon_2)$ (respectively,
$\epsilon_{1c}^{\rm SSS}(\epsilon_2)$) defines the stability threshold of an
incoherent stationary state (ISS) (respectively, the existence
threshold of a synchronized stationary state (SSS)).}
\label{fig:threshold}
\end{figure}

\begin{figure}[!ht]
\includegraphics[width=9cm]{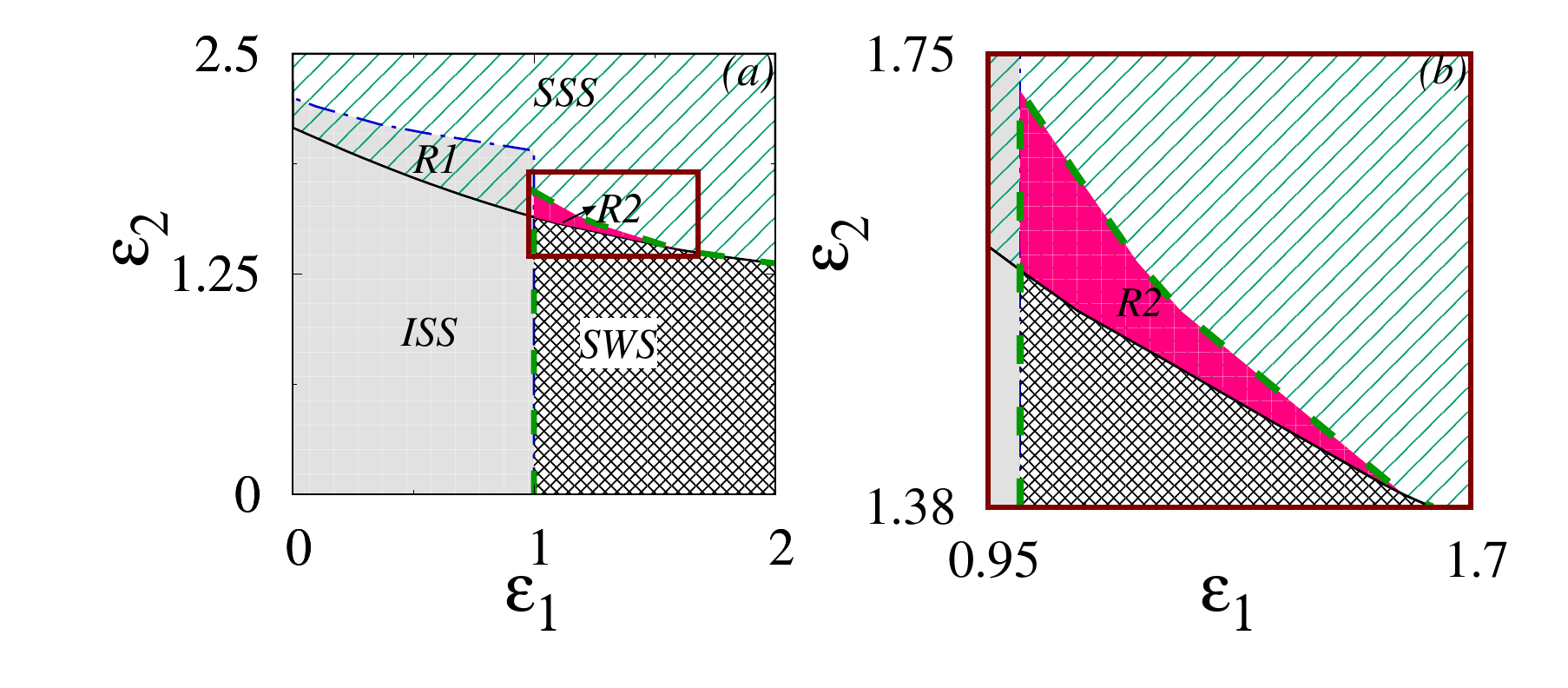}
\caption{For the model~(\ref{eq:eom}) and considering for $\omega_j$'s the Lorentzian distribution~(\ref{eq:lor}) with
$\gamma=0.5$ and $\omega_0=1.0$, the figure depicts the two-parameter bifurcation diagram in the
$(\epsilon_1,\epsilon_2)$-plane. We show here the various stable phases and the phase boundaries, with bifurcation behavior
observed as one crosses the
phase boundaries. (a) The shaded regions, representing
stable existence of the
ISS (Incoherent Stationary State), the SSS (Synchronized Stationary
State) and the SWS (Standing Wave State), have been constructed
        by analyzing the long-time numerical solution of the
        dynamics~(\ref{eq:eom}) for $N=10^5$. 
In numerics, we distinguish between the different regions by requiring
that at long times, $r(t)$ as a function of $t$ behaves differently for the
ISS, the SSS and the SWS. Namely, for the ISS, both the order parameter $r(t)$ and its time average
take the value zero in the $t \to \infty$ limit. In the SWS, $r(t)$ as
$t \to \infty$ oscillates
in time around a time-independent non-zero value, thereby yielding a non-zero time average.
For the SSS, too, $r(t)$ as $t \to \infty$ has a non-zero
time-independent time
average, but it does not oscillate, instead remains equal to a nonzero constant in time. The regions R1 and R2
represent multistability (hysteresis) between ISS-SSS and SSS-SWS,
respectively. The dot-dashed blue line is obtained by using Eq.~(\ref{eq:desyn-critical}), the solid black line is
obtained by using Eq.~(\ref{eq:SS-solution-1}), while the dashed green line is obtained from an analysis of
Eq.~(\ref{eq:OA-r-theta}) by using the numerical package
XPPAUT~\cite{xpp} in which we study the stability of the SWS; Note that all these lines are
        based on the Ott-Ansatz-reduced dynamics corresponding to the
        dynamics~(\ref{eq:eom}). (b)
Enlarged view of the boxed region in (a).} 
\label{fig:phase-diagram}
\end{figure}

We will unravel in this paper a rather rich phase diagram exhibited
by the dynamics~(\ref{eq:eom}). We will show that for a given
$\epsilon_2 \ne 0$, a stationary state (namely, a state for which $z$ is
stationary) occurs only for $\epsilon_1 <
\epsilon_{1c}^{\rm ISS}(\epsilon_2)$ and for $\epsilon_1 >
\epsilon_{1c}^{\rm SSS}(\epsilon_2)$, where $\epsilon_{1c}^{\rm
ISS}(\epsilon_2)$ (respectively,
$\epsilon_{1c}^{\rm SSS}(\epsilon_2)$) defines the stability threshold of an
incoherent stationary state (ISS) (respectively, the existence
threshold of a synchronized stationary state (SSS)). The relative
magnitude of $\epsilon_{1c}^{\rm ISS}(\epsilon_2)$ and
$\epsilon_{1c}^{\rm SSS}(\epsilon_2)$ depends on the value of
$\epsilon_2$, the
relative ordering is reversed. Figure~\ref{fig:threshold} shows, 
based on our analysis presented later in the paper, representative
results for $\epsilon_{1c}^{\rm ISS}(\epsilon_2)$ and
$\epsilon_{1c}^{\rm SSS}(\epsilon_2)$ for a Lorentzian $g(\omega)$,
Eq.~(\ref{eq:lor}).
In Fig.~\ref{fig:phase-diagram}, we show in the
$(\epsilon_1-\epsilon_2)$-plane the various stable phases for the
model~(\ref{eq:eom}) and the phase boundaries. The different phases are
distinguished on the basis of the long-time behavior of the order
parameter $r(t)$. Namely, for the ISS, one has in the long-time ($t \to
\infty$) limit both the order parameter $r(t)$
and its time average, defined as 
\begin{equation}
R\equiv \lim_{t \to \infty} \frac{1}{\tau}\int_t^{t+\tau} {\rm d}t'~r(t'),
\label{eq:R-definition}
\end{equation}
taking the value zero; thus, we have $r_{\rm
st}=0$, and also, $R= 0$. Note that for the ISS, we have ${\rm d}\psi/{\rm
d}t \ne 0$ at long times. In the standing wave state (SWS), $r(t)$ as
$t \to \infty$ oscillates
in time around a time-independent non-zero value. Thus, in this case,
$r(t)$ does not have a stationary value, but we have $R \ne 0$; also,
${\rm d}\psi/{\rm d}t \ne 0$ at
long times. 
For the SSS, too, $R \ne 0$, but $r(t)$ here does not
oscillate, instead remains equal to a non-zero constant in time. For the
SSS, $\psi$ at long times assumes a time-independent value so that
${\rm d}\psi/{\rm d}t=0$ at long times. On the basis of the
foregoing, we summarize that the mean-field frequency $\Omega \equiv {\rm d}\psi/{\rm
d}t$ is zero for the SSS, while the same for the ISS and the SWS have time-dependent values. The regions R1 and R2 in the phase
diagram~\ref{fig:phase-diagram} represent multistability (hysteresis) between ISS-SSS and SSS-SWS,
respectively. At a fixed $\epsilon_1$ and on tuning $\epsilon_2$ (or
vice versa),
one observes phase transitions/bifurcations as one crosses the different phase
boundaries. Figure~\ref{fig:phase-diagram}, the phase diagram or the
bifurcation diagram of model~(\ref{eq:eom}), is the key result of our
work. To summarize the nature of the different displayed phases, we have:
\begin{itemize}
\item ISS: $r_{\rm st}=0,~R=0$. 
\item SWS: $r(t)$ oscillates with time, $r_{\rm st}$ does not exist,
$R \ne 0$. 
\item SSS: $r_{\rm st} \ne 0,~R \ne 0$. 
\end{itemize}
Thus, only the ISS and the SSS qualify as a stationary state (time-independent $z$). With
respect to Fig.~\ref{fig:phase-diagram}, we note that the phase diagram
contains regions of both first-order and continuous transitions.
As
usual, the former is the case whenever on tuning the parameters, one
encounters a region of multistability in going from one phase to another
in the phase diagram, while a continuous transition
happens in the absence of encountering a region of multistability. For $\epsilon_2=0$, when our model~(\ref{eq:eom}) reduces to the
Kuramoto model, one has however only a continuous and no first-order transition.

The rest of the paper is devoted to a derivation of
the aforementioned results.
For Lorentzian $g(\omega)$, Eq.~(\ref{eq:lor}), we use exact
analytical results derived by applying the so-called Ott-Antonsen (OA) ansatz, 
combined with numerical integration of the dynamics~(\ref{eq:eom}) for
large $N$, to
support the bifurcation diagram of Fig.~\ref{fig:phase-diagram}. For the
Gaussian $g(\omega)$, Eq.~(\ref{eq:gaussian}), we present
numerical results to suggest existence of similar stable phases as in the case
of the Lorentzian. The OA ansatz allows
to rewrite in the limit $N \to \infty$ the dynamics of coupled networks of phase oscillators in terms of a few
collective variables~\cite{Ott:2008,Ott:2009}. The power of the ansatz, which explains its
usefulness and its wide applicability, lies in its remarkable ability to
capture precisely and quantitatively, through the dynamical equations for
these collective variables, all, and not just some, of the order parameter attractors and
bifurcations of the dynamics. The latter features may be obtained
directly by performing numerical integration
of the defining equations of motion for $N \gg 1$ and evaluating $r(t)$ in numerics.

The paper is organized as follows. In the next section, we discuss the
OA-ansatz-reduced dynamical equations for the model~(\ref{eq:eom}) for
the choice of the Lorentzian $g(\omega)$, Eq.~(\ref{eq:lor}), while in Section~\ref{sec:OA-results}, we discuss the implications of
the reduced dynamics for the existence of the incoherent and the
synchronized stationary state as well as the standing wave state. In Section~\ref{sec:numerics}, we present
and discuss results obtained from numerical integration of the dynamics~(\ref{eq:eom}) for the Lorentzian and the Gaussian $g(\omega)$,
Eqs.~(\ref{eq:lor}) and~(\ref{eq:gaussian}), respectively, and
compare for the former the numerical and the OA-ansatz-based analytical
results. The paper ends with conclusions in
Section~\ref{sec:conclusions}.

%%%%%%%%%%%%%%%%%%%%%%%%%%%%%%%%%%%%%%%%%%%%%%%%%%%%%%%%%%%%%%%%%%%%%%%%%%%%%%%%%%%%%%%%%%%%
\section{Analysis of the dynamics~(\ref{eq:eom}): The Ott-Antonsen
(OA) ansatz}
\label{sec:analysis}

We now provide an analysis of the dynamics~(\ref{eq:eom}), equivalently,
the dynamics (\ref{eq:eom-1}), in the limit
$N \to \infty$, by invoking the OA ansatz. In this limit, the dynamics (\ref{eq:eom-1}) may be
characterized by the single-oscillator distribution function $f(\theta,\omega,t)$, defined such that $f(\theta,\omega,t){\rm d}\theta$
gives the probability out of oscillators with natural frequency
$\omega$ to find an oscillator with phase in the range
$[\theta,\theta+{\rm d}\theta]$ at time $t$. The distribution is
$2\pi$-periodic in $\theta$ and obeys the normalization 
\begin{equation}
\int_0^{2\pi} {\rm d}\theta~f(\theta,\omega,t)=g(\omega)~\forall~\omega.
\label{eq:norm}
\end{equation}
The
$N\to \infty$ generalization of Eq.~(\ref{eq:r}) reads 
\begin{equation}
	(r_x,r_y)(t)\equiv \int {\rm d}\omega\int_0^{2\pi} {\rm d}\theta~(\cos \theta,\sin
\theta)f(\theta,\omega,t)g(\omega).
\label{eq:r-continuum}
\end{equation}
Since the dynamics (\ref{eq:eom-1}) conserves the number of all
oscillators with a given $\omega$, the time evolution of $f$ follows the
continuity equation
\begin{equation}
\frac{\partial f}{\partial t}+\frac{\partial }{\partial
\theta}\left[\left(\omega+(\epsilon_1+\epsilon_2) r_y
\cos \theta+ (\epsilon_2-\epsilon_1)r_x \sin \theta\right)f\right]=0.
\label{eq:continuity-equation}
\end{equation}
Being $2\pi$-periodic in $\theta$, one may effect a Fourier expansion of
$f$ as
\begin{equation}
f(\theta,\omega,t)=\frac{g(\omega)}{2\pi}\left[1+\sum_{n=1}^\infty
\left(\alpha_n(\omega,t) e^{\im n\theta}+{\rm c.c.}\right)\right],
\label{eq:f-Fourier}
\end{equation}
where the prefactor of $g(\omega)$ ensures that the normalization
(\ref{eq:norm}) is satisfied, $\alpha_n(\omega,t)$ is the $n$-th Fourier
coefficient, while c.c. denotes the term obtained by complex conjugation
of the first term within the brackets. The OA ansatz consists in assuming~\cite{Ott:2008,Ott:2009} 
\begin{equation}
\alpha_n(\omega,t)=\left[\alpha(\omega,t)\right]^n,
\label{eq:OA}
\end{equation}
where the arbitrary function $\alpha(\omega,t)$ is assumed to satisfy
$|\alpha(\omega,t)| < 1$, together with the requirements that
$\alpha(\omega,t)$ may be analytically
continued to the whole of the complex-$\omega$ plane, it has no singularities in the lower-half
complex-$\omega$ plane, and $|\alpha(\omega,t)| \to 0$ as ${\rm
Im}(\omega) \to -\infty$. 

Using the choice~(\ref{eq:OA}) in
Eq.~(\ref{eq:f-Fourier}) defines in the
space ${\cal D}$ of all possible distributions $f(\theta,\omega,t)$ a particular class defined on a manifold ${\cal M}$
in ${\cal D}$. Originally implemented in the context of the
Kuramoto model for a Lorentzian distribution of the
oscillator frequencies, it was shown that this particular class of $f$
remains confined to the manifold ${\cal M}$ under the time
evolution and yields a single first-order ordinary
differential equation for the evolution of the synchronization order
parameter $r(t)$~\cite{Ott:2008,Ott:2009}.

In order to proceed with the OA-ansatz-analysis of the dynamics~(\ref{eq:eom-1}), we consider a Lorentzian $g(\omega)$,
see Eq.~(\ref{eq:lor}). Using Eqs.~(\ref{eq:f-Fourier}) and (\ref{eq:OA}) in
Eq.~(\ref{eq:r-continuum}) yields
\begin{widetext}
\begin{eqnarray}
r_x&=&\frac{1}{2}\int_{-\infty}^\infty {\rm
        d}\omega~g(\omega)\left(\alpha^\star(\omega,t)+\alpha(\omega,t)\right)=\frac{1}{4{\rm
        i}\pi}\oint_C {\rm
d}\omega~\left[\frac{1}{(\omega-\omega_0)-\im\gamma}-\frac{1}{(\omega-\omega_0)+\im\gamma}\right]\left[\alpha^\star(\omega,t)+\alpha(\omega,t)\right]
\label{eq:rx-integral}, \\
r_y&=&\frac{1}{2\im}\int_{-\infty}^\infty {\rm
d}\omega~g(\omega)\left(\alpha^\star(\omega,t)-\alpha(\omega,t)\right)=-\frac{1}{4\pi}\oint_C
{\rm
d}\omega~\left[\frac{1}{(\omega-\omega_0)-\im\gamma}-\frac{1}{(\omega-\omega_0)+\im\gamma}\right]\left[\alpha^\star(\omega,t)-\alpha(\omega,t)\right], \label{eq:ry-integral}
\end{eqnarray}
\end{widetext}
where $\star$ denotes complex conjugation, and where the contour $C$
consists of the ${\rm Re}(\omega)$-axis closed by a large semicircle in
the lower-half complex-$\omega$ plane. In obtaining the second equality
in Eqs.~(\ref{eq:rx-integral}) and (\ref{eq:ry-integral}), we have used
the form~(\ref{eq:lor}) and the fact that the contribution to the
contour integral from the semicircular part of the contour vanishes in view of $|\alpha(\omega,t)|\to 0$ as ${\rm Im}(\omega) \to -\infty$. Evaluating the above integrals by the residue theorem, we get
\begin{eqnarray}
	&&r_x=\frac{\alpha((\omega_0-\im\gamma),t)+\alpha^\star((\omega_0-\im\gamma),t)}{2},\nonumber \\
	\label{eq:order parameter}\\
        &&r_y=\frac{\alpha^\star((\omega_0-\im\gamma),t)-\alpha((\omega_0-\im\gamma),t)}{2{\rm
        i}}.
        \nonumber
\end{eqnarray}
On the other hand, using the expansion (\ref{eq:f-Fourier}) and the
ansatz (\ref{eq:OA}) in Eq.~(\ref{eq:continuity-equation}) and
collecting and equating the coefficient of $e^{i\theta}$ to zero give
\begin{eqnarray}
&&\frac{\partial \alpha(\omega,t)}{\partial t}+\im\omega
\alpha(\omega,t)+\frac{\epsilon_1+ \epsilon_2}{2}\im r_y[1+
\alpha^2(\omega,t)]\nonumber \\
&&+\frac{\epsilon_2- \epsilon_1}{2}r_x[1
-\alpha^2(\omega,t)]=0.
\label{eq:alpha-equation}
\end{eqnarray}
	
Using Eqs.~(\ref{eq:r}) and~(\ref{eq:order parameter}), the Kuramoto order parameter is obtained as
\begin{equation}
z=\alpha^\star((\omega_0-\im\gamma),t).
\label{eq:order parameter1}
\end{equation}
Equation~(\ref{eq:alpha-equation}) then gives
\begin{equation}
\frac{\partial z}{\partial t}-\im(\omega_0+\im\gamma)
z-\frac{\epsilon_1+ \epsilon_2}{2}\im r_y[1+
z^2]+\frac{\epsilon_2- \epsilon_1}{2}r_x[1
-z^2]=0,
\label{eq:alpha-equation1}
\end{equation}
which on using Eq.~(\ref{eq:order parameter}) gives
\begin{equation}
\frac{\partial z}{\partial t}-\im(\omega_0+\im\gamma)z-\frac{\epsilon_1}{2}(z-|z|^2z)+\frac{\epsilon_2}{2}(z^\star-z^3)=0.
\label{eq:alpha-equation2}
\end{equation}
Equation (\ref{eq:alpha-equation2}) rewritten in terms of the quantities $r$ and
$\psi$, see Eq.~(\ref{eq:z}), gives the following two coupled equations:
\begin{eqnarray}
\label{eq:r-dynamics}
\frac{{\rm d}r}{{\rm d}t}&=&-\gamma
r+r(1-r^2)\left(\frac{\epsilon_1}{2}-\frac{\epsilon_2}{2}
\cos(2\psi)\right), \nonumber \\
\label{eq:OA-r-theta} \\ 
\frac{{\rm d}\psi}{{\rm d}t}&=&\omega_0+\frac{\epsilon_2}{2}
(1+r^2)\sin(2\psi). \nonumber
\end{eqnarray}
The above equations constitute the OA-ansatz-reduced order parameter dynamics corresponding to the dynamics~(\ref{eq:eom-1}) in the limit $N
\to \infty$.
Note that for $\epsilon_2=0$, when one has the Kuramoto model, the two
equations in (\ref{eq:OA-r-theta}) are decoupled, and there is only uniform
rotation of $\psi$ with frequency $\omega_0$, that is, the mean-field
frequency equals $\omega_0$; this case was analyzed
in Ref.~\cite{Ott:2008}. For $\epsilon_2 \ne 0$, however, the situation
is much more intricate, as we show below. 

%%%%%%%%%%%%%%%%%%%%%%%%%%%%%%%%%%%%%%%%%%%%%%%%%%%%%%%%%%%%%%%%%%%%%%%%%%%%%%%%%%%%%%%%%%%%
\section{Analysis of the OA-ansatz-reduced dynamics}
\label{sec:OA-results}

\subsection{Incoherent stationary state (ISS):}
\label{subsec:iss}
The dynamics~(\ref{eq:alpha-equation2}), equivalently the dynamics~(\ref{eq:OA-r-theta}), allows for an incoherent stationary state
(ISS) given by $z=z^\star=0$ for all values of $\epsilon_1$ and
$\epsilon_2$; correspondingly, one has $r_{\rm st}=0$, and hence,
$R=0$. The linear stability of this state is determined by linearizing 
Eq. (\ref{eq:alpha-equation2}) around $z=0$, by using the expansion
$z=u$ with $|u|\ll 1$. To this end, we obtain the linear equation 
\begin{equation}
\frac{\partial u}{\partial
t}-\im\left(\omega_0+\im\gamma-\im\frac{\epsilon_1}{2}\right)u+\frac{\epsilon_2}{2}u^\star=0.
\label{eq:alpha-equation3}
\end{equation}
Writing $u=u_x + \im u_y$ yields
\begin{equation}
\frac{\partial }{\partial t}
\begin{bmatrix}
u_x \\
u_y
\end{bmatrix}
=M
\begin{bmatrix}
u_x \\
u_y
\end{bmatrix};
~~M \equiv \begin{bmatrix}
-\gamma+\frac{\epsilon_1}{2}-\frac{\epsilon_2}{2} & -\omega_0 \\
\omega_0& -\gamma+\frac{\epsilon_1}{2}+\frac{\epsilon_2}{2} \\
\end{bmatrix}.
\end{equation}
The matrix $M$ has eigenvalues 
\begin{equation}
\lambda_{1,2}=\frac{-2\gamma +\epsilon_1 \pm
\sqrt{\Delta}}{2},
\end{equation}
with $\Delta
\equiv \epsilon_2^2-4\omega_0^2$. For a given $\epsilon_2$, the
stability threshold for the ISS is then obtained as
\begin{eqnarray}
&&\mathrm{(i)}~~\epsilon_{1c}^{\rm ISS}(\epsilon_2)=2\gamma  \;\;\;
\mbox{for}  \;\;\;\Delta<0, \nonumber \\
\label{eq:desyn-critical} \\
&&\mathrm{(ii)}~~\epsilon_{1c}^{\rm ISS}(\epsilon_2)=2\gamma-\sqrt{\Delta}   \;\;\; \mbox{for}
\;\;\; \Delta>0. \nonumber
\end{eqnarray}

\subsection{Synchronized stationary state (SSS):}
\label{subsec:sss}
Considering the dynamics~(\ref{eq:OA-r-theta}), we now explore the
possibility of existence of a synchronized stationary state (SSS), i.e.,
$r_{\rm st}\ne 0$, and hence, $R\ne 0$. In the
case of the Kuramoto model, this would mean to have in the laboratory
frame a state with
time-independent $r$ together with $\psi$ changing uniformly in time with
frequency $\omega_0$; in this case, on transforming to a frame rotating uniformly with
frequency $\omega_0$ with respect to the laboratory frame, one obtains
the SSS in which both $\psi$ and $r$ and, hence, $z$ have time-independent
values. Considering the dynamics~(\ref{eq:OA-r-theta}) and requiring $r$
and $\psi$ to have time-independent non-zero
values $(r_{\rm st},\psi_{\rm st})$ so that the left hand side of the two equations
in~(\ref{eq:OA-r-theta}) may be set to zero, we obtain for
the SSS the two coupled equations
\begin{eqnarray}
\frac{\epsilon_2}{2}\cos(2\psi_{\rm st})&=&\frac{\gamma}{
(r_{\rm st}^2-1)}+\frac{\epsilon_1}{2}, \nonumber \\
\label{eq:OA-r-theta-SSS} \\ 
\frac{\epsilon_2}{2}\sin(2\psi_{\rm st})&=&\frac{-\omega_0}
{(1+r_{\rm st}^2)}. \nonumber
\end{eqnarray}
The above equations yield the following solutions for $(r_{\rm
st},\psi_{\rm st})$:
\begin{eqnarray}
\left(\frac{\gamma }{r_{\rm
st}^2-1}+\frac{\epsilon_1}{2}\right)^2+\left(\frac{\omega_0}{1+r_{\rm
st}^2}\right)^2=\frac{\epsilon_2^2}{4}, \label{eq:SS-solution-1}\\
\tan(2\psi_{\rm st})=\frac{\omega_0 (1-r_{\rm st}^2)}
{(1+r_{\rm st}^2)(\gamma+\frac{\epsilon_1}{2}(r_{\rm
st}^2-1))}.\label{eq:SS-solution-2}
\end{eqnarray}
The first equation implies that for a given $\epsilon_1$, no real $r_{\rm
st}$ value, and hence, no SSS exist for $\epsilon_2=0$. In fact, for a given $\epsilon_1$, an SSS
exists for $\epsilon_2$ larger than a critical value
$\epsilon_{2c}^{\rm SSS}\equiv \epsilon_{2c}^{\rm SSS}(\epsilon_1)$. Alternatively, for a
given $\epsilon_2$, there exists a critical $\epsilon_{1c}^{\rm SSS}(\epsilon_2)$
beyond which the SSS exists.

\subsection{Standing Wave State (SWS):}
\label{subsec:dss}
A standing wave state (SWS) is characterized by the order
parameter $r(t)$ at long times oscillating as a function of $t$, but
nevertheless yielding a non-zero time average at long times, $R \ne 0$.
It is thus distinct from a synchronized stationary state (SSS) for which
both the order parameter and its time average have a non-zero value at
long times, but the former does not oscillate as a function of time. 
Deriving stability conditions for the SWS does not prove easy, unlike
the ISS and the SSS. Hence, we analyzed using Eq.~(\ref{eq:r-dynamics})
the SWS stability by employing the numerical package XPPAUT~\cite{xpp}. The period of the SWS is obtained by solving the time-dependent equations~(\ref{eq:OA-r-theta}); we were however unable to obtain an analytical solution of the said equations.

In the next section, we view the above results \textit{vis-\`{a}-vis}
results obtained from numerical integration of the
dynamics~(\ref{eq:eom}) for large $N$.

%%%%%%%%%%%%%%%%%%%%%%%%%%%%%%%%%%%%%%%%%%%%%%%%%%%%%%%%%%%%%%%%%%%%%%%%%%%%%%%%%%%%%%%%%%%%
\section{Numerical results}
\label{sec:numerics}

\begin{figure}	
\begin{center}
\includegraphics[width=8cm]{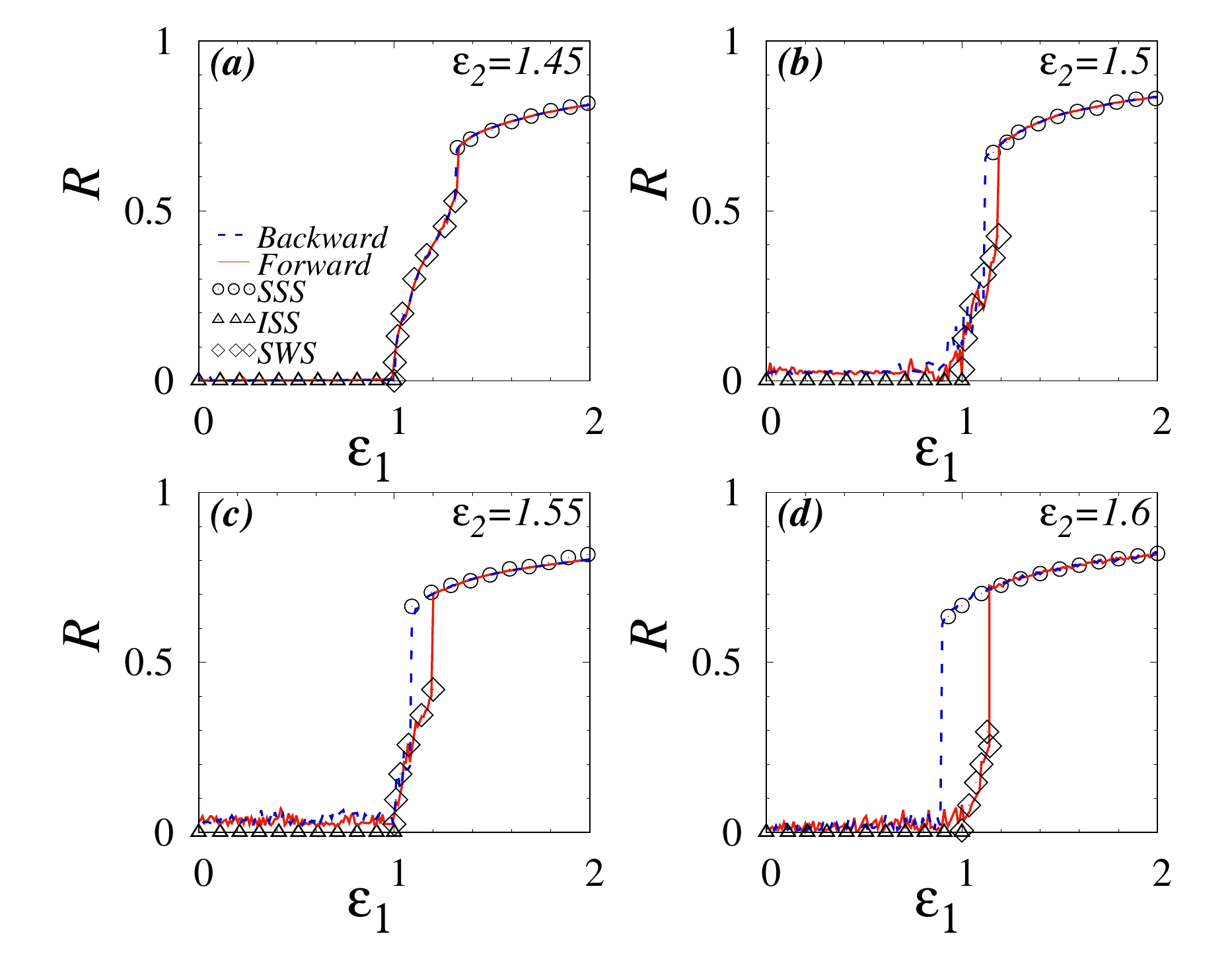}
\includegraphics[width=8cm]{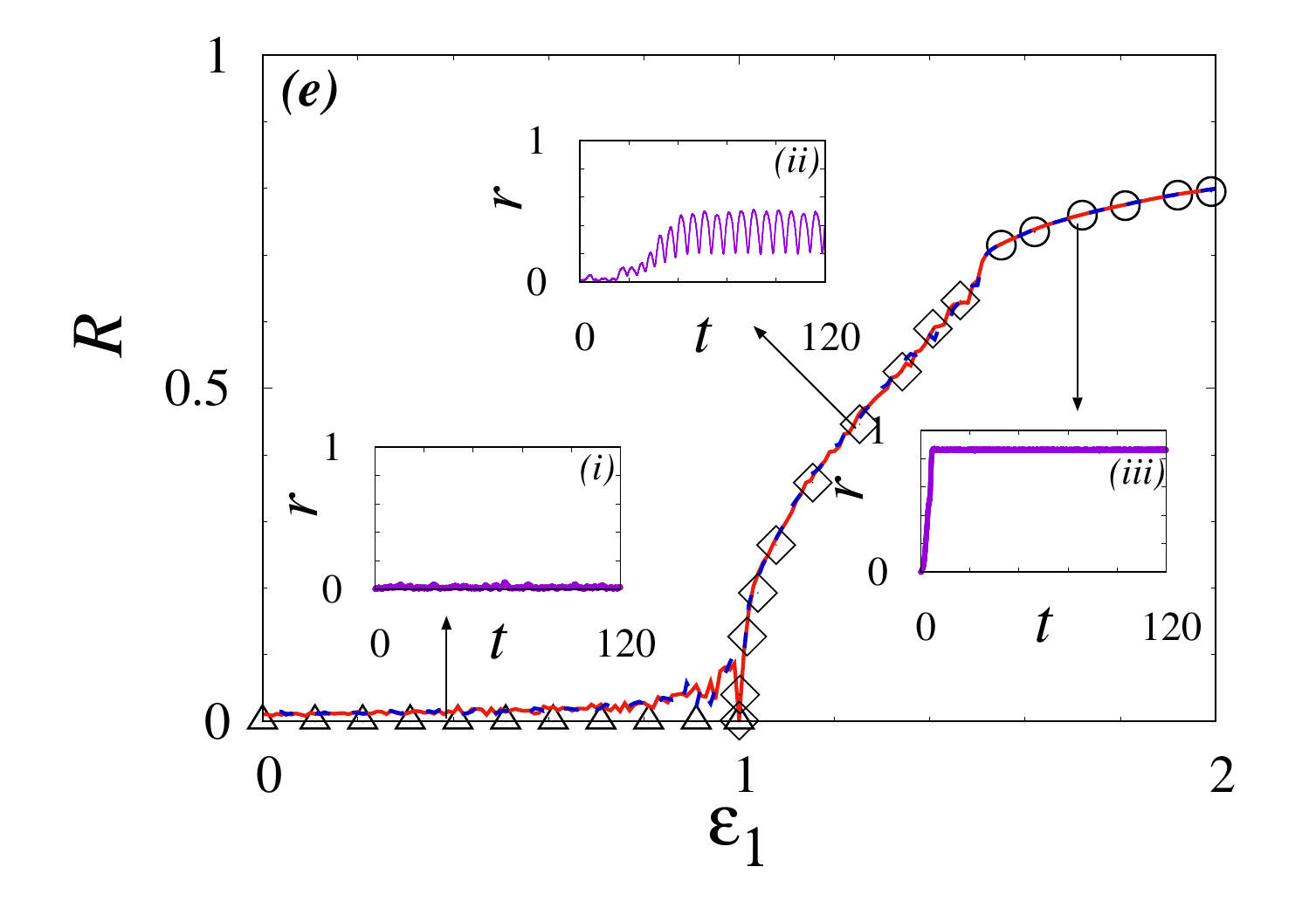}
\caption{The first four panels show the quantity $R$, namely, the time-averaged order parameter in the long-time limit,
see Eq.~(\ref{eq:R-definition}), as a function of adiabatically-tuned $\epsilon_1$ and for four
different values of $\epsilon_2$, see text.  Panel (e) shows $r(t)$ as a function of $t$ for $\epsilon_2=1.4$ and for
three representative $\epsilon_1$
values, namely, (i) $\epsilon_1=0.35$, when we have an ISS (see
Fig.~\ref{fig:phase-diagram}), (ii) $\epsilon_1=1.2$, when we have a SWS,
and (iii) $\epsilon_1= 1.7$, when we have a SSS. In all cases, the lines are obtained from numerical
integration of the dynamics~(\ref{eq:eom}) for $N=5\times 10^5$,
while symbols correspond to predictions based on the OA-ansatz-reduced
dynamics~(\ref{eq:OA-r-theta}) discussed in Section~\ref{sec:OA-results}. The frequency distribution is a
Lorentzian, see Eq.~(\ref{eq:lor}), with $\gamma=0.5$ and $\omega=1.0$.}
\label{fig:2-3}
\end{center}
\end{figure}

\begin{figure}	
\begin{center}
\includegraphics[width=8cm]{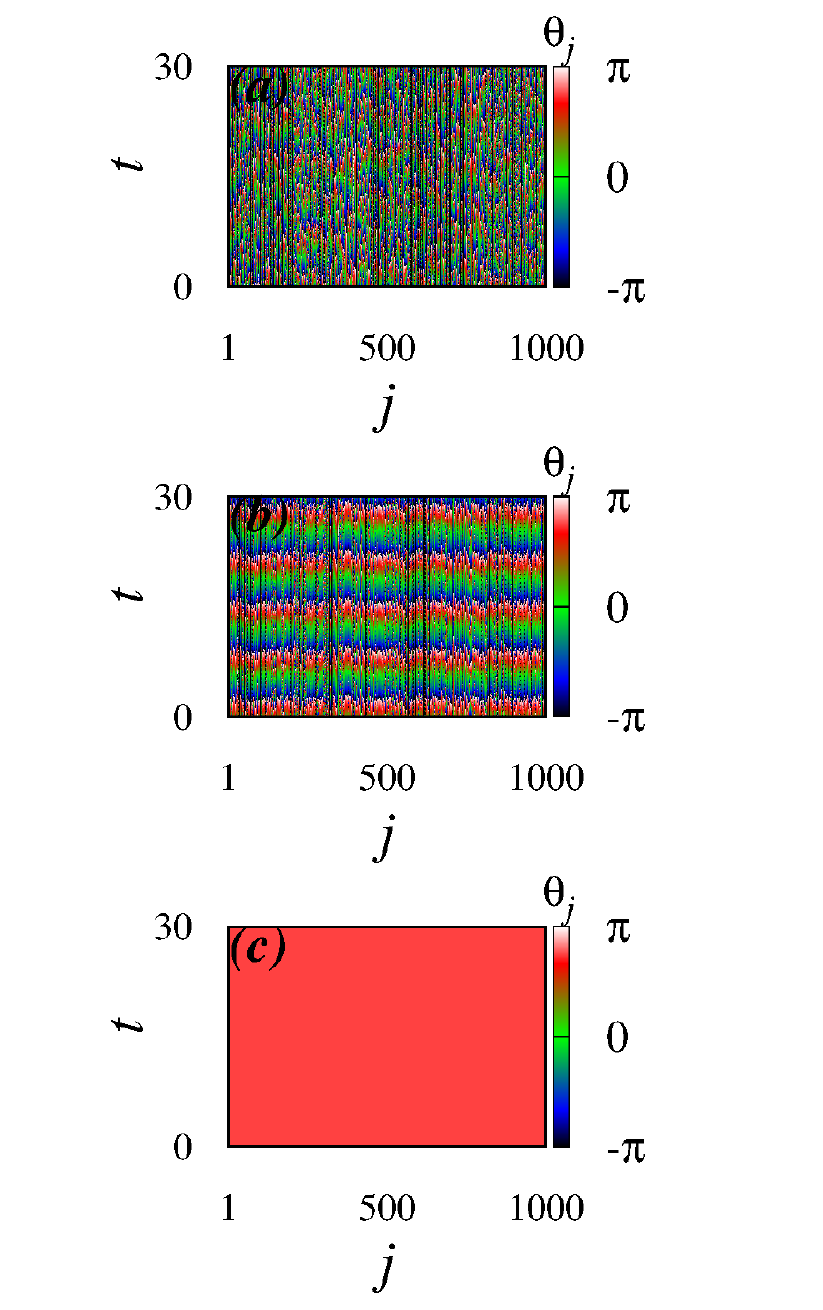}
        \caption{Corresponding to the insets in panel (e) of Fig.~\ref{fig:2-3},
        the figure shows $\theta_j$ for different $j$ and as a
        function of time at long times (the figure depicts the results not for the full range
        of $j$ but for $j=1,2,\ldots,1000$). The color coding given on the
        side of each panel denotes the intensity of $\theta$
        values. The top panel corresponds to the
        ISS, the middle panel to the SWS and the bottom panel to the
        SSS. As may be seen from the figure, unlike the ISS and the SSS,
        the SWS exhibits a stationary wave pattern: at a fixed $j$, the
        value of $\theta$ changes periodically as a function of
        time. The data are obtained 
        from numerical integration of the dynamics~(\ref{eq:eom}) with
        parameter values and other details same as in
        Fig.~\ref{fig:2-3}(e).}
\label{fig:space-time}
\end{center}
\end{figure}

\begin{figure}	
\begin{center}
\includegraphics[width=8cm]{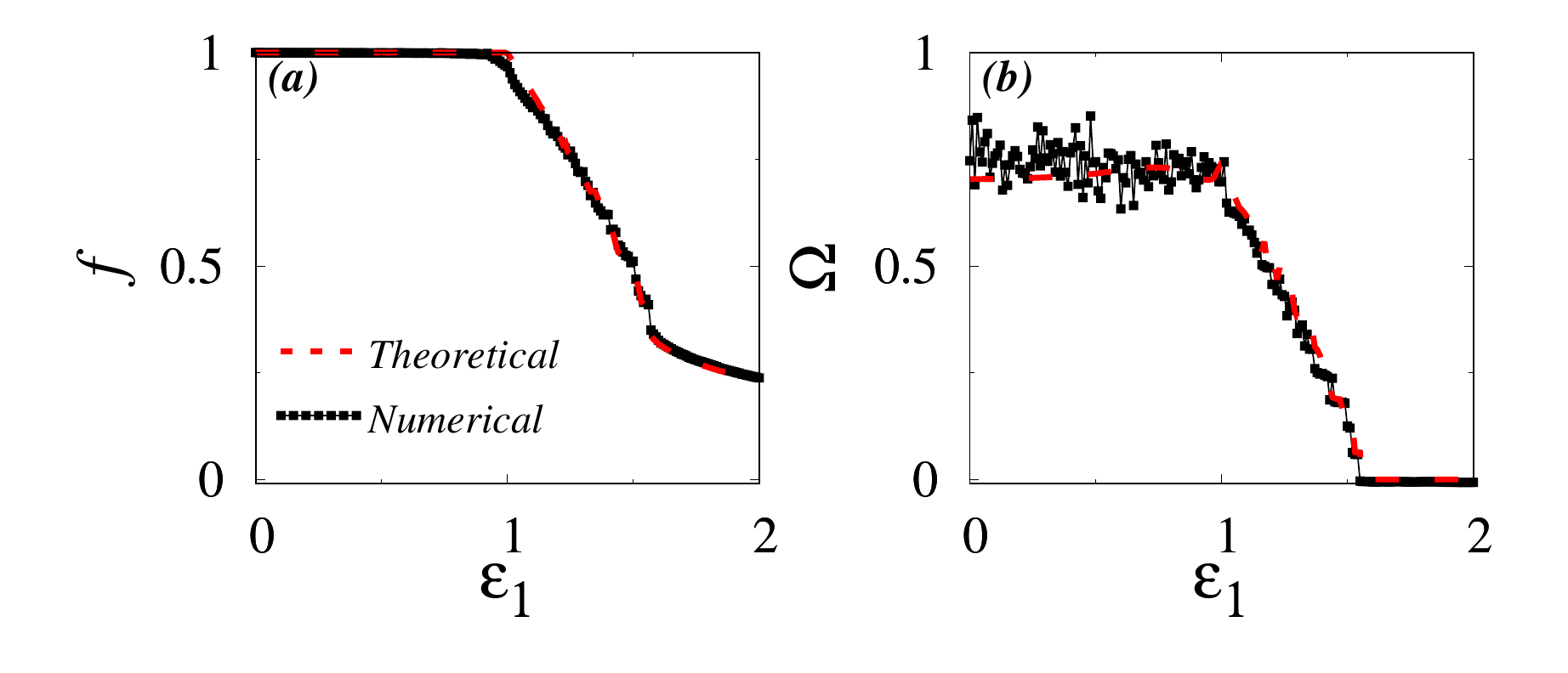}
        \caption{Corresponding to the panel (e) in Fig.~\ref{fig:2-3},
        the figure shows as a function of $\epsilon_1$ the
        mean-ensemble frequency $f$ and the mean-field frequency
        $\Omega$ at long times. Here, we have compared numerical
        integration results with those based on the OA ansatz; for
        details of computation, see Section~\ref{sec:numerics}.} 
\label{fig:f}
\end{center}
\end{figure}

\begin{figure}	
\begin{center}
\includegraphics[width=8cm]{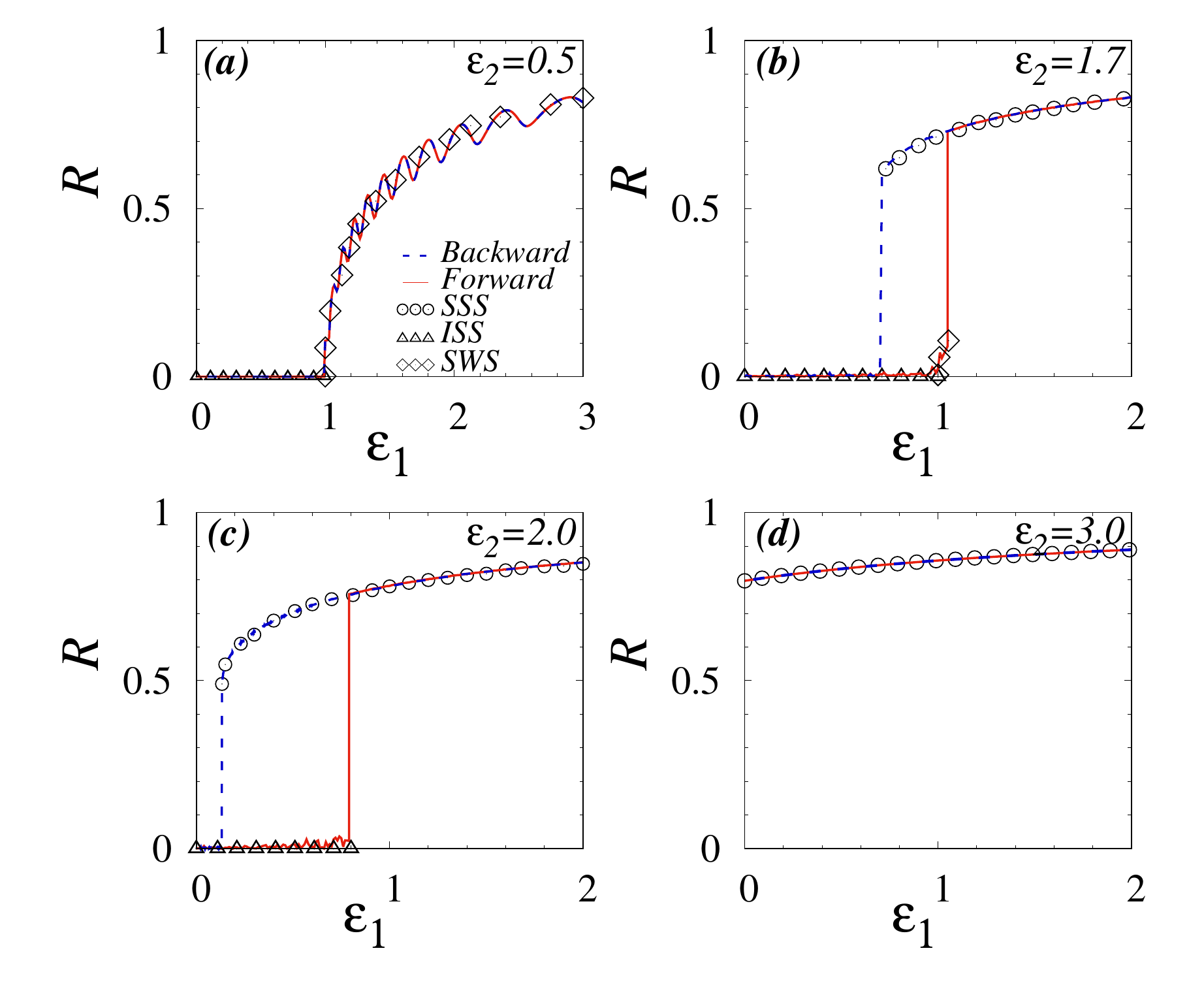}
\caption{The four panels show the quantity $R$, namely, the time-averaged order parameter in the long-time limit,
see Eq.~(\ref{eq:R-definition}), as a function of adiabatically-tuned $\epsilon_1$ and for four
different values of $\epsilon_2$, see text. In all cases, the lines are obtained from numerical
integration of the dynamics~(\ref{eq:eom}) for $N=5\times 10^5$,
while symbols correspond to predictions based on the OA-ansatz-reduced
dynamics~(\ref{eq:OA-r-theta}) discussed in Section~\ref{sec:OA-results}. The frequency distribution is a
Lorentzian, see Eq.~(\ref{eq:lor}), with $\gamma=0.5$ and $\omega=1.0$,}
\label{fig:4}
\end{center}
\end{figure}

\begin{figure}	
\begin{center}
\includegraphics[width=8cm]{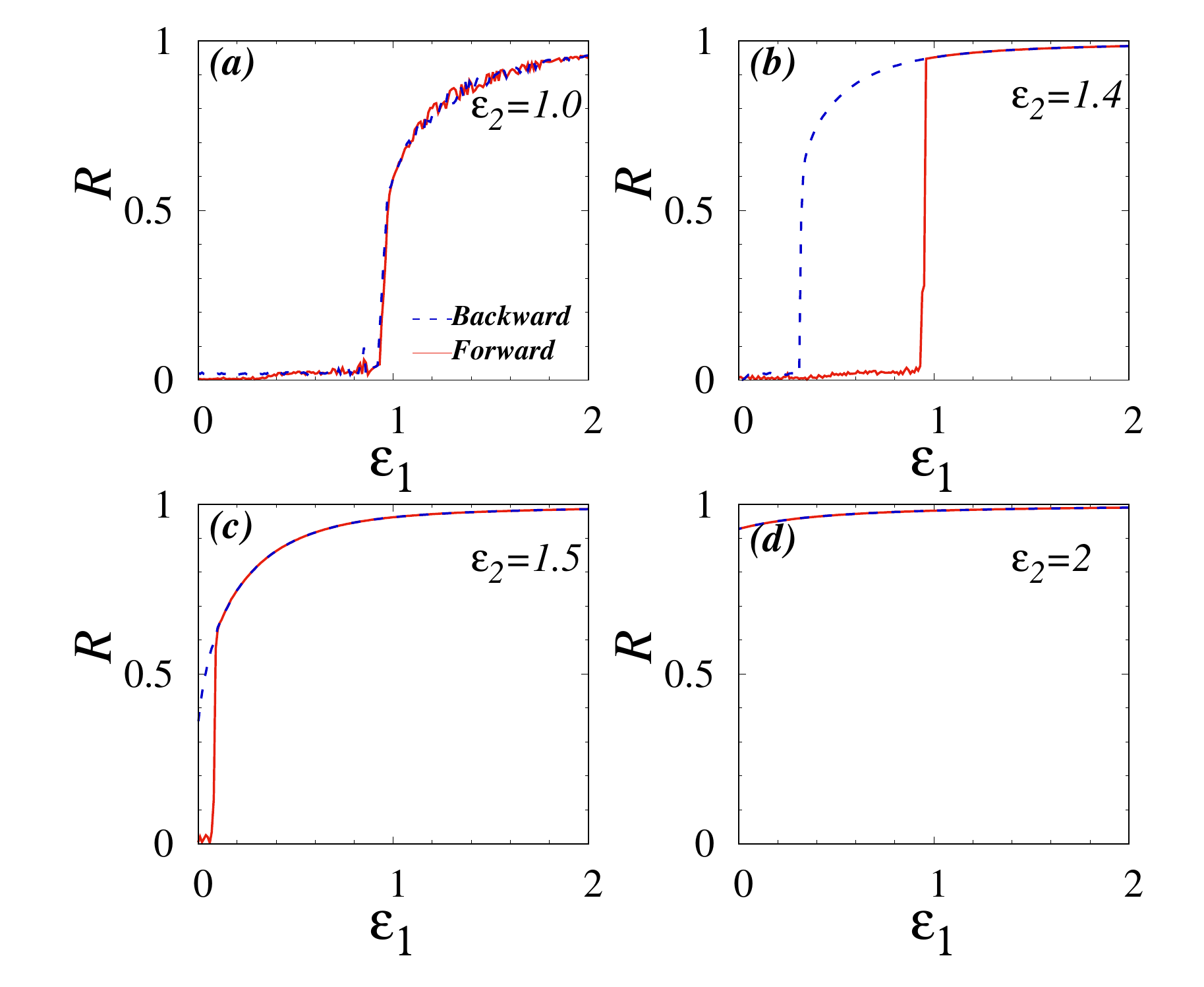}
\caption{The four panels show the quantity $R$, namely, the time-averaged order parameter $R(t)$ in the long-time limit,
see Eq.~(\ref{eq:R-definition}), as a function of adiabatically-tuned $\epsilon_1$ and for four
different values of $\epsilon_2$, see text. In all cases, the lines are obtained from numerical
integration of the dynamics~(\ref{eq:eom}) for $N=5\times 10^5$. The frequency distribution is a
Gaussian, see Eq.~(\ref{eq:gaussian}), with $\omega_0=1.0$ and
$\sigma=0.5$.}
\label{fig:6}
\end{center}
\end{figure}

We start off this section with a discussion of results based on the OA-ansatz-reduced dynamics discussed in Section~\ref{sec:OA-results} and on direct numerical integration of the dynamics~(\ref{eq:eom}), in the case of the Lorentzian
distribution~(\ref{eq:lor}) for the frequencies. The numerical
integration involved use of a standard fourth-order Runge-Kutta
integration algorithm with integration time step ${\rm d}t=0.01$. Within the OA-reduced dynamics, the analytical results we should be focusing on are 
Eq.~(\ref{eq:desyn-critical}) for the stability threshold
$\epsilon_{1c}^{\rm ISS}(\epsilon_2)$ of the ISS and
Eqs.~(\ref{eq:SS-solution-1}) and~(\ref{eq:SS-solution-2}), with the
latter yielding the
existence threshold $\epsilon_{1c}^{\rm SSS}(\epsilon_2)$ of the SSS, as
we now detail. Choosing $\gamma=0.5$ and $\omega_0=1.0$ in Eq.~(\ref{eq:lor}), we may obtain
$\epsilon_{1c}^{\rm SSS}(\epsilon_2)$ for a fixed $\epsilon_2$ by varying
$\epsilon_1$, from low to high values, using
Eq.~(\ref{eq:SS-solution-1}) to record the particular value of
$\epsilon_1$ when for the first time the equation gives a solution for
$r_{\rm st}$ in the range $0 < r_{\rm st} \le 1$, and identifying this particular
value of $\epsilon_1$ with $\epsilon_{1c}^{\rm SSS}(\epsilon_2)$. In Fig.~\ref{fig:threshold}, we
show as a function of $\epsilon_2$ both the thresholds
$\epsilon_{1c}^{\rm ISS}(\epsilon_2)$ and $\epsilon_{1c}^{\rm
SSS}(\epsilon_2)$. 

In Fig.~\ref{fig:2-3}, we show results based on numerical integration of
the dynamics~(\ref{eq:eom}) with $N=5\times 10^5$, for $R$ as a
function of adiabatically-tuned $\epsilon_1$ for various
values of $\epsilon_2$. We first let the system settle to the stationary
state at a fixed value of $\epsilon_1 \approx 0$, and then tune
$\epsilon_1$ adiabatically in time from low to high values while recording the value
of $R$ in time; this corresponds to forward variation of $\epsilon_1$.
Subsequently, we tune $\epsilon_1$ adiabatically in time from high to
low values (backward variation of $\epsilon_1$). Adiabatic tuning ensures that the system remains in the
stationary state at all times as the value of $\epsilon_1$ changes in
time. In all panels in Fig.~\ref{fig:2-3}, we see that the system
exists in either the ISS, the SWS or the SSS. For panel (a), the system
exhibits a continuous transition between the ISS and the SWS, and between
the SWS and the SSS. For panels (b), (c), and (d), the transition between the
ISS and the SWS is still continuous, while the one between the SWS and
the SSS becomes of first order. For $\epsilon_2=1.4$ case, we shows  
 the variation of $R$ as a
function of adiabatically-tuned $\epsilon_1$ via numerical integration
of the dynamics~(\ref{eq:eom}) in Fig.~\ref{fig:2-3}(e); here, the
three insets show for three representative
values of $\epsilon_1$ the variation of $r(t)$ with time at long times.
That the SWS is actually a standing wave is
        evident from the spatiotemporal plots in
        Fig.~\ref{fig:space-time}, where we show at long times the time
        series of $\theta_j$ for different $j$ in a system of size
        $N=10^5$ (the figure depicts the results not for the full range
        of $j$ but for $j=1,2,\ldots,1000$); the parameter values are the same as in
        Fig.~\ref{fig:2-3}(e).
        Here, the color coding refers to the intensity of $\theta$
        values. It is evident from the figure that the SWS indeed
        behaves as a wave stationary in space, while the ISS and the SSS
        do not qualify as a wave.

        Corresponding to Fig.~\ref{fig:2-3}(e), Fig.~\ref{fig:f} shows as a function of $\epsilon_1$ the
        mean-ensemble frequency $f$ and the mean-field frequency
        $\Omega$ at long times. The numerical data plotted in
        the figure correspond to the following. First, one obtains $r_x$ and $r_y$ at long times
        from numerical integration of the dynamics~(\ref{eq:eom}) with
        parameter values and other details same as in
        Fig.~\ref{fig:2-3}(e). The values so obtained for $r_x$ and
        $r_y$ are then averaged over a
        stretch of time interval at
        long times; $f$ is then obtained as
        $f=\omega_0+2\epsilon_2 r_xr_y$. On the other hand, one obtains
        $\Omega$ numerically by substituting the numerically-obtained
        time-averaged values of $r$ and $\psi$ in Eq.~(\ref{eq:OA-r-theta}). The
        figure also shows theoretical data for both $f$ and $\Omega$,
        which are obtained as
        follows: (i) for the ISS, we have $r_x=r_y=0$, giving
        $f=\omega_0$; (ii) for the SWS, we solve numerically
        Eq.~(\ref{eq:OA-r-theta}) to obtain $r(t)$ and $\psi(t)$, from which
        we obtain $r_x(t)$ and $r_y(t)$ as $r_x(t)=r(t)\cos
        \psi(t),~r_y(t)=r(t)\sin
        \psi(t)$. We then estimate $f$ from values of $r_x(t)$ and
        $r_y(t)$ averaged over time at long times; (iii) for the SSS, we obtain $r$ and $\psi$
        at long times as being equal to the time-independent values
        $r_{\rm st}$ and $\psi_{\rm st}$, respectively, obtained by
        solving numerically Eq.~(\ref{eq:OA-r-theta-SSS}). For $\Omega$,
        we have: (i) For the ISS, $\Omega$ is obtained by solving
        numerically for the
        time-dependent $\psi$ the
        second equation in~(\ref{eq:OA-r-theta}) with $r=0$; (ii) For
        the SWS, $\Omega$ is obtained by solving numerically for $r$ and
        $\psi$ as a function of $t$ the coupled
        equations~(\ref{eq:OA-r-theta}), and evaluating $\Omega$ as
        $\Omega={\rm d}\psi/{\rm d}t$. (iii) For the SSS, we
        have $\Omega=0$. We see from the figure a very good match
        between numerical integration results and results based on the
        OA ansatz. For a discussion on the different behavior of
        mean-field and mean-ensemble frequencies in the context of the
        Kuramoto model, we refer the reader to
        Ref.~\cite{Petkoski}.

In Fig.~\ref{fig:4}, we show for four values of $\epsilon_2$ the
variation of $R$ with $\epsilon_1$ adiabatically tuned over a
wider range than is considered in Fig.~\ref{fig:2-3}. We see (i) in
panel (a), for which $\epsilon_{1c}^{\rm ISS}(\epsilon_2)$ is finite and $\epsilon_{1c}^{\rm
SSS}(\epsilon_2) \to \infty$ (see Fig.~\ref{fig:threshold}), no existence
of the SSS and a continuous transition between the ISS and
the SWS, (ii) in panel (b) the transition between the ISS and SWS is continuous, while the transition between the SWS and the SSS is first order, (iii) in panel (c) a first-order transition between the ISS and the SSS, and (iv) in panel (d), for which $\epsilon_{1c}^{\rm ISS}(\epsilon_2)$ is finite and $\epsilon_{1c}^{\rm
SSS}(\epsilon_2) \to 0$, the existence of only the SSS. In all the panels
in Figs.~\ref{fig:2-3} and~\ref{fig:4}, we see a good match of the data
for $R$ obtained from numerical integration of the dynamics~(\ref{eq:eom}) and from the OA-ansatz-reduced-dynamics
discussed in Section~\ref{sec:OA-results}. We have also checked the
match between the
OA-based results for $\epsilon_{1c}^{\rm ISS}(\epsilon_2)$ and
$\epsilon_{1c}^{\rm SSS}(\epsilon_2)$ and those estimated from numerical
integration of the dynamics~(\ref{eq:eom}). Our work thus provides further credence to the validity and the usefulness of the OA-ansatz in describing order
parameter dynamics of globally-coupled phase oscillators, and is an
useful addition to the ever-growing list of references demonstrating the
applicability of the OA approach (Refs.~\cite{Ott1,Ott2,Ott3,Ott4} provide a random sampling of papers on
applications of the OA approach).

For the Gaussian frequency distribution, Eq.~(\ref{eq:gaussian}) with
$\omega_0=1.0,\sigma=0.5$, Fig.~\ref{fig:6} shows results for $R$ as a function of adiabatically-tuned $\epsilon_1$ for four values
of $\epsilon_2$, obtained from numerical integration of the
dynamics~(\ref{eq:eom}) with $N=5 \times 10^5$. We see qualitatively similar
phases as for the Lorentzian case, namely, the ISS, the SWS and the SSS,
with (i) a continuous transition between the ISS and the SWS and between
the SWS and the SSS in panel (a) a first-order transition between the
ISS and th SSS in panels (b) and (c), and (iii) the existence of only the SSS in panel (d).

%%%%%%%%%%%%%%%%%%%%%%%%%%%%%%%%%%%%%%%%%%%%%%%%%%%%%%%%%%%%%%%%%%%%%%%%%%%%%%%%%%%%%%%%%%%
\section{Conclusions}
\label{sec:conclusions}
In this work, we studied a nontrivial generalization of the
celebrated Kuramoto model of spontaneous collective synchronization, by
considering an additional interaction in the dynamics that breaks the
rotational symmetry of the model. The Kuramoto model comprises
limit-cycle oscillators of distributed natural frequencies that are
coupled all-to-all. With the help of direct numerical
integration of the dynamics and exact analytical results based on the
so-called Ott-Antonsen ansatz for the specific case of a Lorentzian
frequency distribution, we unraveled a
rather rich phase diagram of the generalized model
\textit{vis-\`{a}-vis} the
Kuramoto model. The phase diagram contains in it both stationary and standing wave
phases. In the former, the synchronization order parameter $r$ has a
long-time value that is time independent. On the other hand, one has in
the standing wave phase an
oscillatory behavior of the order parameter as a function of time that
nevertheless yields a non-zero and time-independent time average. It would be interesting to study the effect
of rotational-symmetry-breaking interaction on the inertial version of
the Kuramoto model~\cite{Gupta:2018}. Introducing inertia drastically modifies the phase
diagram of the Kuramoto model, so we may already anticipate new
features on adding the symmetry-breaking interaction. Investigations in
this direction are under way and will be reported elsewhere.

%%%%%%%%%%%%%%%%%%%%%%%%%%%%%%%%%%%%%%%%%%%%%%%%%%%%%%%%%%%%%%%%%%%%%%%%%%%%%%%%%%%%%%%%%%%
\section{Acknowledgements}
The work of V.K.C. is supported by the SERB-DST-MATRICS Grant No.
MTR/2018/000676 and  CSIR Project under Grant No. 03(1444)/18/EMR-II.
M.M. wishes to thank SASTRA Deemed University for research funds and extending infrastructure support to carry out this work. S.G. acknowledges support from the Science
and Engineering Research Board (SERB), India under SERB-TARE scheme Grant No.
TAR/2018/000023 and SERB-MATRICS scheme Grant No. MTR/2019/000560. He also thanks ICTP -- The Abdus Salam International Centre for Theoretical Physics,
Trieste, Italy for support under its Regular Associateship scheme. 

%%%%%%%%%%%%%%%%%%%%%%%%%%%%%%%%%%%%%%%%%%%%%%%%%%%%%%%%%%%%%%%%%%%%%%%%%%%%%%%%%%%%%%%%%%%
\appendix

\section{Motivating the form of the dynamics~(\ref{eq:eom})}
\label{app1}

Here, we motivate the form of the dynamics~(\ref{eq:eom}). To this end, let us consider a collection of $N$ globally-coupled Stuart-Landau
limit-cycle oscillators with conjugate feedback, with dynamics given by
\begin{equation}
\frac{{\rm d}z_j}{{\rm d}t}=(1+\im\omega_j)z_j-|z_j|^2z_j+\frac{1}{N}\left[\epsilon_1
\sum_{k=1}^N (z_k-z_j)-\epsilon_2\sum_{k=1}^N z_k^*\right],
\label{eq:sl1}
\end{equation}
where the complex number $z_j$ characterizes the $j$-th oscillator,
$j=1,2,\ldots,N$, the quantity $\epsilon_1$ denotes the strength of a
diffusive coupling between the oscillators, while $\epsilon_2$ is the
mean-field feedback strength. Writing $z_j$ in terms of real quantities
$R_j;~0 \le R_j \le 1$, and $\theta_j \in [-\pi,\pi]$, as
$z_j=R_je^{\im\theta_j}$, Eq.~(\ref{eq:sl1}) gives
\begin{eqnarray}
&&\frac{{\rm d}R_j}{{\rm d}t}=(1-\epsilon_1-R_j^2)R_j\nonumber \\
&&+\frac{1}{N}\left[\epsilon_1
\sum_{k=1}^N R_k\cos{(\theta_k-\theta_j)}-\epsilon_2\sum_{k=1}^N R_k
\cos{(\theta_k+\theta_j)}\right], \nonumber \\ \label{eq:LS-1}\\
&&\frac{{\rm d}\theta_j}{{\rm d}t}=\omega_j\nonumber \\
&&+\frac{1}{N}\left[\epsilon_1
\sum_{k=1}^N
\frac{R_k}{R_j}\sin{(\theta_k-\theta_j)}+\epsilon_2\sum_{k=1}^N
\frac{R_k}{R_j}\sin{(\theta_k+\theta_j)}\right]. \nonumber \\ \label{eq:LS-2}
\label{eq:sl2}
\end{eqnarray}
In order to analyze the above dynamics, let us first consider the
noninteracting case: $\epsilon_1=\epsilon_2=0$. It is then easily checked that
Eq.~(\ref{eq:LS-1}) has fixed points $R_j=R_{\rm s}\equiv
 1~\forall~j$ and $R_j=R_{\rm u}\equiv
0~\forall~j$, of which the former is stable and the latter is unstable.
The long-time dynamics then corresponds to a limit-cycle for each of the
individual oscillators with corresponding frequency
$\omega_j$ and amplitude equal to $R_{\rm s}$, and with
the corresponding motion described by the phase-only dynamics~${\rm
d}\theta_j/{\rm d}t=\omega_j~\forall~j$. When the couplings
$\epsilon_1$ and $\epsilon_2$ are sufficiently weak, a perturbation
theory about the aforementioned limit-cycle behavior would imply
substituting $R_j=R_{\rm s}~\forall~j$ in Eq.~(\ref{eq:LS-2}), thus reducing it to a set
of $N$ coupled equations of the form (\ref{eq:eom}). 

%%%%%%%%%%%%%%%%%%%%%%%%%%%%%%%%%%%%%%%%%%%%%%%%%%%%%%%%%%%%%%%%%%%%%%%%%%%%%%%%%%%%%%%%%%%


\begin{thebibliography}{99}
\bibitem{Pikovsky:2001}A. Pikovsky, M. Rosenblum and J. Kurths, \textit{Synchronization: a Universal Concept in Nonlinear
Sciences} (Cambridge University Press, Cambridge, 2001).
\bibitem{Buck:1988}J. Buck, {\it Synchronous rhythmic flashing of
fireflies. II.}, Q. Rev. Biol. {\bf 63}, 265 (1988).
\bibitem{Peskin:1975}C. S. Peskin, {\it Mathematical aspects of heart
physiology} (Courant Institute of Mathematical Sciences, New York,
1975).
\bibitem{Kiss:2002}I. Kiss, Y. Zhai and J. Hudson, {\it Emerging coherence in a population of chemical oscillators}, Science {\bf
296}, 1676 (2002).
\bibitem{Temirbayev:2012}A. A. Temirbayev, Z. Zh. Zhanabaev, S. B. Tarasov, V. I. Ponomarenko and
M. Rosenblum, {\it Experiments on oscillator ensembles with global
nonlinear coupling}, Phys. Rev. E {\bf 85}, 015204(R) (2012).
\bibitem{Benz:1991}S. P. Benz and C. J. Burroughs, {\it Coherent
emission from two‐dimensional Josephson junction arrays}, Appl. Phys. Lett. {\bf 58}, 2162 (1991).
\bibitem{Zeda:2000}Z. N\'{e}da, E. Ravasz,
T. Vicsek, Y. Brechet and A. L. Barab\'{a}si, \textit{Physics of the
rhythmic applause}, Phys. Rev. E {\bf 61},
6987 (2000). 
\bibitem{Rohden:2012}M. Rohden, A. Sorge, M. Timme and D. Witthaut, {\it
Self-Organized synchronization in decentralized power grids}, Phys. Rev.
Lett. {\bf 109}, 064101 (2012).
\bibitem{Scholl}R. Berner, J. Sawicki and E. Sch\"{o}ll, \textit{Birth
and Stabilization of Phase Clusters by Multiplexing of Adaptive
Networks}, Phys. Rev. Lett. {\bf 124}, 088301 (2020).
\bibitem{Kuramoto:1984}Y. Kuramoto, \textit{Chemical Oscillations, Waves
and Turbulence} (Springer, Berlin, 1984).
\bibitem{Strogatz:2000}S. H. Strogatz, \textit{From Kuramoto to Crawford: exploring the onset of synchronization in
populations of coupled oscillators}, Physica D {\bf 143}, 1 (2000).
\bibitem{Acebron:2005}J. A. Acebron, L. L. Bonilla, C. J. P. Vicente, F.
Ritort and R. Spigler, \textit{The Kuramoto model: a
simple paradigm for synchronization phenomena}, Rev. Mod. Phys. {\bf
77}, 137 (2005).
\bibitem{Gupta:2014}S. Gupta, A. Campa and S. Ruffo, \textit{Kuramoto model of synchronization: equilibrium and
nonequilibrium aspects}, J. Stat. Mech. R08001 (2014).
\bibitem{Gupta:2018}S. Gupta, A. Campa and S. Ruffo, \textit{Statistical
Physics of Synchronization} (Springer, Berlin, 2018).
\bibitem{Martens:2009}E. A. Martens, E. Barreto, S. H. Strogatz, E. Ott,
P. So and T. M. Antonsen, \textit{Exact results for the Kuramoto model with a bimodal frequency distribution}, Phys. Rev. E {\bf 79}, 026204 (2009).
\bibitem{Aneta1}D. Iatsenko, S. Petkoski, P. V. E. McClintock and A.
Stefanovska, \textit{Stationary and Traveling Wave States of the
Kuramoto Model with an Arbitrary Distribution of Frequencies and
Coupling Strengths}, Phys. Rev. Lett. {\bf 110}, 064101 (2013).
\bibitem{Aneta2}D. Iatsenko, P.V.E. McClintock and A. Stefanovska,
\textit{Glassy states and super-relaxation in populations of coupled
phase oscillators}, Nature Communications {\bf 5}, 4118 (2014).
\bibitem{xpp}B. Ermentrout, \textit{Simulating, Analyzing, and Animating Dynamical Systems: A Guide
	to XPPAUT for Researchers and Students } (Society for Industrial \& Applied Math,
Philadelphia, PA, 2002).
\bibitem{Ott:2008}E. Ott and T. M. Antonsen, \textit{Low dimensional
behavior of large systems of globally coupled oscillators}, Chaos {\bf 18}, 037113
(2008).
\bibitem{Ott:2009}E. Ott and T. M. Antonsen, \textit{Long time evolution of phase oscillator systems}, Chaos {\bf 19}, 023117
(2009).
\bibitem{Petkoski}S. Petkoski, D. Iatsenko, L. Basnarkov, and A.
        Stefanovska, \textit{Mean-field and mean-ensemble frequencies of
        a system of coupled oscillators}, Phys. Rev. E {\bf 87}, 032908
        (2013).
\bibitem{Ott1}O. E. Omel'chenko, \textit{Partially coherent twisted states in arrays of coupled phase oscillators}, M. Wolfrum and C. Laing, Chaos {\bf
24}, 023102 (2014).
\bibitem{Ott2}C. R. Laing, \textit{Traveling waves in arrays of delay-coupled phase oscillators}, Chaos {\bf 26}, 094802 (2016).
\bibitem{Ott3}J. G. Restrepo and P. S. Skardal, \textit{Competitive suppression of synchronization and nonmonotonic transitions
in oscillator communities with distributed time delay}, Phys. Rev. Research {\bf 1}, 033042
(2019).
\bibitem{Ott4}O E Omel'chenko, \textit{Traveling chimera states in systems of phase oscillators with asymmetric nonlocal coupling}, Nonlinearity {\bf 33}, 611 (2020).
\end{thebibliography}
\end{document}